\documentclass[12pt,a4papper]{article}

\usepackage{graphics}
\usepackage{graphicx}
\usepackage{amssymb}
\usepackage{latexsym}
\usepackage{amsmath}
\usepackage{amsfonts}
\usepackage{color}

\newcommand{\bfbeta}{\mbox{\boldmath$\beta$}}
\newcommand{\beq}{\begin{equation}}
\newcommand{\eeq}{\end{equation}}
\newcommand{\dif}{{\rm d}}

\begin{document}

\title{On Mach's principle: Inertia as gravitation}
\author{J. Mart\'in\thanks{Departamento de F\'{\i}sica
Te\'orica, Universidad de Salamanca, Salamanca, Spain, e-mail:
chmm@usal.es}\and Antonio F. Ra\~{n}ada\thanks{Facultad de
F\'{\i}sica, Universidad Complutense, 28040 Madrid, Spain, e-mail:
afr@fis.ucm.es}\and  A. Tiemblo\thanks{Instituto de Matem\'aticas y
F\'{\i}sica Fundamental, CSIC, Serrano 113b, 28006 Madrid, Spain,
e-mail: tiemblo@imaff.cfmac.csic.es}}
\date{21 March 2007}
\maketitle

\begin{abstract}
In order to test the validity of Mach's principle, we calculate the
action of the entire universe on a test mass in its rest frame,
which is an acceleration ${\bf g}^*$. We show the dependence of the
inertia principle on the lapse and the shift. Using the formalism of
linearized gravitation, we obtain the non-relativistic limit of
${\bf g}^*$ in terms of two integrals. We follow then two
approaches.  In the first one, these integrals are calculated in the
actual time section $t=t_0$ up to the distance $R_U=ct_0$. In the
more exact and satisfactory second approach, they are calculated
over the past light cone using the formalism of the retarded
potentials. The aim is to find whether the acceleration $\dot{\bf
v}$ in the LHS of Newton's second law can be interpreted as a
reactive acceleration, in other words, as minus the acceleration of
gravity ${\bf g}^*$ in the rest frame of the accelerated particle
({\it i. e.} to know whether or not ${\bf g}^*=-\dot{\bf v}$). The
results strongly support Mach's idea since the reactive acceleration
for $\Omega _\Lambda =0.7$ turns out to be about ${\bf
g}^*=-1.1\,\dot{\bf v}$, in the first approach, and about ${\bf
g}^*= -0.7\,\dot{\bf v}$, in the second. These results depend little
on $\Omega_\Lambda$ if $\Omega_\Lambda<0.9$. Even considering the
approximations and idealizations made during the calculations, we
deem these results as interesting and encouraging.
\end{abstract}

\newpage

\tableofcontents

\bigskip

\section{Introduction}

\subsection{Mach's critique of Newton's laws}

Elaborating further the ideas put forth by the physicist
Christiaan Huygens and the philosophers Gottfried Leibniz and
Bishop Berkeley, Mach proposed a radical criticism of Newton's
absolute space, more than thirty years before Einstein's first
paper on relativity \cite{Mac72,Mac83}.
 He said in 1872 ``For me only relative motion exists ...
When a body rotates relative to the fixed stars, centrifugal
forces are produced; when it rotates [but] not relative to the
fixed stars, no centrifugal forces are produced ... Obviously it
does not matter if we think of the earth as turning round on its
axis, or at rest while the fixed stars revolve round it". From
this premise, he concluded that the inertia is not a primary
property of the bodies, as  assumed by Newton in his {\it
Principia}, but, quite on the contrary, an effect of its motion
with respect to the fixed stars (or the distant galaxies, in
current parlance). In other words, inertia would be an interaction
that requires other bodies to manifest itself, so that it would
have no sense in a universe consisting of only one mass.

Mach's criticism had an influence on Einstein, who coined the
expression ``Mach principle" to denote the idea that inertia is an
interaction with all the mass in the universe, to be expressed
eventually by mathematical relations still to be discovered.
 Indeed  Einstein himself
acknowledged Mach's influence in several occasions. In his student
days, he had read with attention his main work \cite{Mac83}, ``a
book which, with its critical attitudes toward basic concepts and
basic laws, made a deep and lasting impression on me" \cite{See60}.
When Mach died in 1916, he wrote in an obituary ``... It is not
improbable that Mach would have discovered the theory of relativity,
if, at the time when his mind was still young and susceptible, the
problem of the constancy of the speed of light had been discussed
among physicists" \cite{Ein16} and in a letter to A. Weiner in 1930,
``it is justified to consider Mach as the precursor of the general
theory of relativity" \cite{Pai82}. Later, however, Einstein's
interest in Mach's work waned, when he came to think that ``Mach was
a good scientist but a wretched philosopher", a comment certainly
due to his negative attitude towards the atoms. Nevertheless as Pais
says, ``After Einstein, the Mach principle faded but never died".

Indeed, it is usually accepted that there must be something
important in Mach's principle, in particular that it has some
relation with general relativity although in a subtle and not yet
understood way. Indeed it is very vague: although Mach was probably
thinking in gravitation, he suggested no explicit mechanism that
could transmit any interaction from a fixed star to a mass. Nobody
was able later to formulate the idea in a working way, so that, in
Berry's words, ``the principle is half-baked" \cite{Ber89}. Sciama
wrote an important and intriguing paper in 1953, suggesting that
inertia and gravitation are the same phenomenon
\cite{Sci53}-\cite{Sci71}. His proposal was certainly appealing, if
only qualitative. The aim of this work is to find a way to
understand Mach principle, baking it further in the frame of
linearized general relativity.

\section{Inertia}

According to Mach, this seems clear, the inertia of a test mass $m$
with velocity $\bf v$ and acceleration ${\bf a} =\dot{{\bf v}}$
(overdot means time derivative) adopts the form of a reactive force
${\bf F}_{\rm I}=-m{\bf a}$ due to the fixed stars. Today we must
speak instead not only of the distant galaxies, but of all the
mass-energy in the entire visible universe.  Since he insisted that
only the relative situation with respect to other bodies could be a
cause of the forces, the principle can be given an operative status
as follows.  The reactive force ${\bf F} _{\rm I}$ can be calculated
in the rest frame of the test mass as the gravitational pull of  all
the galaxies moving with acceleration $-\dot{\bf v}$ and velocity
$-{\bf v}$. If this pull were shown to produce the reactive
acceleration \begin{equation} {\bf g}^*=-\dot{{\bf
v}}\,,\label{10}\end{equation}  in the limit $v \ll c$, the
principle would be fully baked, since it would have an operative
character (we will write ${\bf g}^*=-\xi \dot{\bf v}$ in the
following). This work proposes a way to do that in the frame of
linearized general relativity.

This reactive acceleration describes probably one of the most
fundamental properties of the universe. The spacetime being a
dynamical structure, it seems logic from the assumptions of
Newtonian physics to consider inertia as a property of the matter
itself. Nevertheless, on the grounds of General Relativity the
problem adopts a new feature as long as spacetime play the role of a
natural ingredient of that dynamics. Even more, the absence of an
explanation of inertia can be considered as a test for the
completeness of a dynamical theory.

In a general way, the motion of a particle depends entirely on
geometry through the equations of geodesics.
\begin{equation} {{\rm d} ^2 x^\alpha \over {\rm d} \tau ^2}+\Gamma
^{\alpha}_{\;\;\mu\nu}\,{{\rm d} x^\mu\over {\rm d} \tau}\,{{\rm d}
x^\nu\over {\rm d} \tau}=0\, , \label{90}\end{equation} where $\tau$
is the proper time and $\Gamma^\alpha _{\;\,\mu\nu}$ are the
Christoffel symbols, which in the weak field approximation
$g_{\mu\nu}=\eta _{\mu\nu}+h_{\mu\nu}$ are given by
\begin{equation} \Gamma ^{\alpha}_{\;\;\mu\nu} =\frac{1}{2} \, \eta
^{\alpha\gamma}\, (h_{\mu\gamma,\,\nu}+ h_{\nu\gamma,\,\mu}-
h_{\mu\nu,\, \gamma})\, .\label{100}\end{equation} In the Newtonian
limit, the terms with $\mu=\nu =0$ give the dominant part of the
second term in the LHS of (\ref{90}), which can be approximated as $
\ddot{x}^{\,i} =-\Gamma ^{i}_{\;\;00}\,\dot{x}^0\dot{x}^0$, so that
the acceleration of the test mass is \begin{equation} g^*_i =-\Gamma
^i_{\;00}c^2 \, .\label{110}\end{equation}

In the previous equations and in the rest of this work, the
Newtonian three-velocities and three-accelerations, such as $v_i,
g^*_j,\ldots $ with Latin subindexes are the non-relativistic limit
of the relativistic quantities with the same superindexes and that
$\beta _k=v_k/c$.
  The corresponding  Christoffel symbols  and
acceleration of the point mass $g^*_i$ are therefore
\begin{equation} \Gamma ^i_{\;\;00}= -\frac{1}{2} \, (2h_{0i,\,
0}-\, \, h_{00,\, i})\,,\quad g^*_i =c^2 (h_{0i,\,0}-{1\over 2}\,
h_{00,\, i})\, ,\label{120}\end{equation} in the Newtonian
approximation where $\gamma =1$. Strictly speaking,  formula
(\ref{120}) is a prediction for the value of the reactive
acceleration of a particle in geodesic motion, the inertia in other
words.

Even being an unrealistic point of view, the most obvious approach
to get a first insight is to consider the case of pure gravity or,
equivalently, to investigate what geometry itself could suggest us
on the origin of inertia. At this point, a better understanding is
reached by using the formalism of Hamiltonian gravity. For this
purpose, we pass from the usual four-dimensional description to the
classical ADM(3+1) splitting of spacetime, with dynamical variables
$N$, lapse, $N_i$, shift, and $q_{ij}$, three dimensional metric
tensor \cite{DeW64,Alv89}. Therefore, the reactive acceleration
${\bf g}^*$ becomes a function of $N$ and $N_i$.

Dropping out total divergences and time derivatives, the (3+1)
Einstein-Hilbert Lagrangian becomes
\begin{equation} {\cal L}= \int
N\sqrt{q}\,(k_{ij}k^{ij}-k^2+R^{(3)})\,\dif ^3x\,.
\label{EinHil}\end{equation}Here, the latin indexes run over the
three dimensional space, $R^{(3)}$ is the three-dimensional scalar
curvature and
$$k_{ij}={1\over 2N}\,\left( D_{(i}N_{j)}-{\partial q_{ij}\over
\partial t}\right),$$
stands for the extrinsic curvature. There are no time derivatives of
lapse and shift in the Lagrangian density, so that their canonical
momenta vanish supplying two primary constraints. After a brief
calculation, we get the corresponding Hamiltonian in the form
$$H=\int \dif ^3x\,(NS+N_iV^i),$$
where $S$ and $Vî$ are the scalar and vector constraints,
respectively, depending on the metric $q_{ij}$ and its canonical
conjugate momentum $\Pi^{ij}$. The stability of the primary
constraints leads us to the secondary ones $S=0$ and $V^i=0$. It is
easy to verify that all of them ar first class (symmetries), in such
a way that the values of $N$ and $N_i$, and consequently of $g_i^*$,
 can be chosen arbitrarily (gauge fixing, remind that we are in pure gravity).

 Taking, however, into account the commonly accepted universal
 validity of the inertia principle, one is naturally lead to assume
 the existence of a  cosmological universal source able to fix,
  universally as well, the value of $N$ and $N_i$. These arguments
 indicate that something more than geometry is needed, a
 point of view very close to Mach' ideas, in which inertia plays more the
 role of a footprint of the contents of the universe rather
  than a test of General Relativity.

  The next step for an understanding of inertia is, obviously, to add
  particles to the geometry.  One can verify, using Hamiltonian methods, that this addition
  suffices to fix the values of $N$ and $N_i$. In particular, the standard
  model of the universe with dust, radiation and dark energy
   acquires therefore naturally inertia as a
  property.

\section{The cosmological  model and the calculation method}
This paper accepts the standard cosmological model with flat space
sections, {\it i.e.} with critical density $\rho _{\rm cr}(t)$ and
$k=0$ in Friedmann's equation. It assumes also that the universe
consists of matter and dark energy, the latter being equivalent to a
positive cosmological constant $\Lambda$ and the former, either
ordinary or dark, with zero pressure (it can be treated as dust).
The present densities of matter and dark energy are, respectively,
$\rho _{\rm M}=\Omega _{\rm M}\rho _{\rm cr}$ and $\rho _\Lambda =
\Lambda/8\pi G=\rho _{\rm cr}\Omega _\Lambda$, with $\Omega _{\rm
M}+\Omega _\Lambda =1$.

  To show the validity
of Mach's idea  that inertia is not an intrinsic property of
particles and bodies but an interaction with the entire universe
and, furthermore, that this interaction is just gravity, we follow
several steps. (i) We take a point test mass moving with velocity
${\bf v}$ and acceleration $\dot{{\bf v}}$. (ii) We compute the
force of a particular galaxy on this particle in the particle rest
frame, in which the galaxy has opposite velocity $-{\bf v}$ and and
acceleration $-\dot{{\bf v}}$. This is done by solving the particle
equation of motion in the gravitational field of the galaxy, using
the formalism of linearized general relativity and taking the
Newtonian limit with small velocity $v \ll c$. (iii) The sum of the
forces of all the galaxies (plus the intergalactic mass-energy) is
approximated by the integral over the universe of the effect of a
uniform distribution of matter and energy with the critical density
$\rho =\rho _{\rm cr}$. This is possible because nearby bodies as
the Milky Way have a negligible effect as compared with the distant
galaxies.

Note that the linear approximation is used here to calculate the
gravitational field of a galaxy at a point, a standard problem in
celestial mechanics in order to determine the reactive force. This,
of course, does not imply that the metric of the standard model can
be approximated as a Minkowskian one perturbed with a term
proportional to $G$.

The calculations are made for several values of $\Omega _\Lambda$
following two approaches. In the first, the necessary integrals are
evaluated in the actual time section $t=t_0$ up to the distance
$ct_0$, with constant critical density $\rho _{\rm cr}$ and $\Omega
_\Lambda$; in other words, neglecting the time retard. The result
will be that the reactive acceleration is in the interval
$(-1.1\dot{\bf v},\; -1.2\dot{\bf v})$ if  $0\leq \Omega _\Lambda
\leq 0.9$.

The second approach is more exact and satisfactory since the
integrals are calculated over our (onion shaped) past light cone,
taking into account the time retard, for which a singularity has to
be eliminated. We will obtain thus reactive accelerations in the
interval $(-0.7\,\dot{\bf v},\;-0.6\dot{\bf v})$ for $0\leq \Omega
_\Lambda\leq 0.9$.  Being close to (\ref{10}), these are certainly
remarkable results.

In fact as we have pointed out, the existence of inertia means, as a
first requisite, the need to guess the correct dynamics able to fix
the value of $N$ and $N_i$. Even more, the reactive acceleration
must acquire, at least in the dominant terms, an expression of the
form $-\xi\,\dot{\bf v}$, $\xi$ being a constant with a value near
to one. None of these conditions is, by any means, obvious and,
curiously, a universe with particles has precisely these properties.

\section{Linearized general relativity}

\subsection{Field equations in linearized general relativity} In
this work, we will use the linearized equations of general
relativity \cite{Mis73,Rin01}. Assuming the linear approximation
to gravity with $g_{\mu\nu} =\eta _{\mu\nu}+h_{\mu\nu}$, $\eta
_{\mu\nu}\equiv \mbox{diag} (1, \, -1,\, -1,\,  -1)$,
$|h_{\mu\nu}|\ll 1$, Einstein's equations take the approximate
form \begin{eqnarray} &&\qquad \qquad{\huge \Box} h_{\mu\nu}
={16\pi G\over c^4}\, \left(T_{\mu\nu}-\frac{1}{2} \eta
_{\mu\nu}T\right)\, ,\nonumber\\&&\hspace{10cm}
\label{20}\\&&\qquad \qquad \partial _\mu (h_\nu^{\;\;\mu}-{1\over
2}\,\delta_\nu^\mu h)=0\,,\nonumber
\end{eqnarray} where $h=\eta ^{\alpha\beta}h_{\alpha\beta}$,
$T_{\mu\nu}$ is the energy-momentum tensor and $T$ its trace.

The solution of (\ref{20}) can be found by means of the retarded
Green function as
\begin{equation} h_{\mu\nu}(x)= {16\pi
G\over c^4}\, \int D_{\rm r}(x-x^\prime )\left(T_{\mu\nu}
(x^\prime)-{1\over 2}\,\eta _{\mu\nu}T (x^\prime )\right)\, {\rm
d} ^4 x^\prime \,\label{30}\end{equation} where
\begin{equation} D_{\rm r} (x-x^\prime) = {1\over 2\pi}\,\theta
(x_0-x^\prime _0)\, \delta [(x-x^\prime )^2]\,
,\label{40}\end{equation} $ \theta$ being the step or Heaviside
function (see, for instance,  \cite{Jac98}).

 In the cases of a perfect fluid
without pressure (as in a dust universe) and of a point mass $M$,
the energy-momentum tensor takes the forms, respectively,
\begin{equation} T_{\mu\nu} =\rho u_\mu u_\nu\,, \qquad T_{\mu\nu}
(x)= Mc\int {\rm d} \tau \, u_\mu (\tau) u_\nu (\tau)\delta
^{(4)}[x-r(\tau )]\,, \label{50}\end{equation} where $\rho$ is the
mass density, $u_\mu ={\rm d} x_\mu /{\rm d} \tau$ the
four-velocity field, $\tau$ the proper time and $r (\tau)$ the
four-trajectory of the mass.

\subsection{The effect of a galaxy}

Equation (\ref{20}) for the gravitational field created by a point
galaxy of mass $M$ at the position of the test particle in its
rest frame can be solved with well known standard methods
\cite{Jac98}. The galaxy has three-velocity $\bf v$ and
three-acceleration $\dot{\bf v}$ and its energy-momentum tensor
$T^{\mu\nu}$ is given by the second equation (\ref{50}). We assume
that the space-time positions of the galaxy and the test particle
are $r(\tau)$ and $x$, $\tau$ being the proper time of the galaxy,
so that the four-vector from the galaxy to the particle is
$x-r(\tau )$. Note that ${\rm d} {\bf R}/{\rm d}\tau =- {\bf v}$
and ${\rm d}R/{\rm d}\tau =-{\bf n}\cdot {\bf v}$.

The solution of (\ref{20}) is
\begin{equation} h_{\mu\nu}(x)= -{4GM\over c^4}\,c\,
\left.{u_\mu (\tau) u_\nu (\tau )-\eta _{\mu\nu}c^2/2\over u\cdot
[x-r(\tau )]}\right|_{\rm ret}\, ,\label{60}\end{equation} where the
subindex means that  the RHS must be evaluated at the retarded time
$\tau ^\prime$.  A comment is in order here. Eq. (\ref{60}) does not
verify ``exactly" the second eq. (\ref{20}); in fact this divergence
turns out to be proportional to the product of the gravitational
constant $G$ times the four-acceleration of the galaxy at the
retarded point. Now, the linear approximation that we are carrying
out is the first order part of a formal expansion in series of
powers of $G$ \cite{Bel81}, which is tantamount to say in powers of
non-dimensional parameters depending on the masses of the problem.
Consequently, both $h_{\mu\nu}$ and the four-acceleration are
proportional to $G$. In other words, this method is correct: the
first equation (\ref{20}) shows that the acceleration is of order
$G$, so that the second is verified at linear order.

\section{First approach to the reactive acceleration}

 Once known
the expressions of $h_{\alpha\beta}$ from (\ref{60}) and taking into
account (\ref{90})-(\ref{110}), we recover the expression
(\ref{120}) to calculate ${\bf g}^*$ from (\ref{60}). Following then
a standard method, it is easy to find in the limit of small velocity
$\beta \ll 1$, that
\begin{equation} h_{0i,\,0}=(4GM/c^4)\dot{{\bf v}}/R\,, \qquad
h_{00,\,i}=(2GM/c^3)[(\dot{\bfbeta}\cdot{\bf n})n_i/R+n_ic/R^2]\,,
\label{130}\end{equation} where $\bf R$ is the three-vector from
the test mass to the galaxy, $\dot{\bf v}$ is the galaxy
acceleration and ${\bf n}={\bf R}/R$. All the quantities are taken
at the retarded time. In this first approach we start with a dust
universe of constant density $\rho _{\rm cr}(t_0)$ consisting in
all the galaxies plus the intergalactic matter, which produce a
gravitational field like (\ref{130}). The acceleration of the test
mass is then obtained by integration in the time section until the
distance $ct_0$.

To obtain the acceleration of the test mass, we insert (\ref{130})
into (\ref{120}), change the sign of $\dot{\bf v}$ and integrate
over the time section $t=t_0$ until the radius $R_U=ct_0$ with the
critical density $\rho _{\rm cr}=3H_0^2/8\pi G$. This leads to
\begin{equation}{\bf g}^*=-{16\pi G\over c^2}\,\dot{{\bf v}}
\int _0^{R_U}\rho_{\rm cr}R{\rm d}R  + {4\pi G\over 3c^2}\,
\dot{\bf v}\,\int _0^{R_U}\rho _{\rm cr} R{\rm d}R\,, \label{140}
\end{equation}
where the two terms are the contribution to the reactive
acceleration of the terms in $h_{0i,\,0}$ and $h_{00,\,i}$,
respectively. Since we take constant $\rho_{\rm cr}$, the
integrals are immediate. It is easy to see that the effect of
terms containing $h_{ij}$ or $h_{ii}$ can be neglected in the
limit $\beta \rightarrow 0$.

Three points must be underscored: i) we are accepting the
approximation that all the galaxies have the same acceleration; ii)
the terms proportional to $\bf v$ are not considered in (\ref{130})
because we are taking the limit $\beta \ll 1$; and iii) it is easy
to see that the terms in $h_{ij}$ or $h_{ii}$ can be neglected in
the same limit.

\paragraph{The dust universe.} If $(\Omega _{\rm M},\Omega _\Lambda )= (1,\,0)$, it happens
that $H_0t_0=2/3$, the total reactive acceleration on a test mass
being therefore
\begin{equation} {\bf g}^*=-{11\over 4}\,(H_0t_0)^2\dot{{\bf v}}=-{11\over 9}\,\dot{{\bf v}}=-1.22\, \dot{{\bf v}}\,
.\label{150}\end{equation} Being close to the correct value
$-\dot{\bf v}$, this result is remarkable. However, we have not yet
considered the dark energy.

\subsection{Introducing the dark energy}
 In the case of a perfect fluid
with mass density $\rho$ and pressure $p$, the expression of the
energy-momentum tensor is
\begin{equation} T_{\mu\nu} =(\rho +p/c^2) u_\mu u_\nu\ -\eta
_{\mu\nu}p\,. \label{70}\end{equation} As is well known, both the
dark energy and the cosmological constant have negative pressure. In
fact $\rho =\rho_{\rm cr}(t_0)=\rho _{\rm M}+\rho _\Lambda$ is the
addition of the mass densities of matter and dark energy, and $p=
p_\Lambda =-\rho _\Lambda\, c^2$ is the pressure of the dark energy.
Since $\rho _\Lambda =\Omega _\Lambda \rho_{\rm cr}$, $T_{\mu\nu}$
and its trace can be written
\begin{equation}  T_{\mu\nu}=\rho_{\rm cr} (1-\Omega _\Lambda )u_\mu
u_\nu+\rho _{\rm cr}c^2\,\Omega _\Lambda \eta _{\mu\nu} \,,\quad
T=\rho_{\rm cr}  u_\mu u^\mu +p u_\mu u^\mu /c^2 -4p = \rho_{\rm cr}
c^2-3p\, .\label{80}\end{equation}

\paragraph{The standard model.} The source of $h_{\mu\nu}$ in (\ref{20}) is $T_{\mu\nu}-\frac{1}{2} \eta
_{\mu\nu}T$, so that those of $h_{00}$ and $h_{0i}$ are,
respectively,
\begin{equation} \rho _{\rm cr}(u_0)^2(1-\Omega _\Lambda )-\frac{1}{2}
\rho_{\rm cr} c^2(1+\Omega_\Lambda)\quad  \mbox{ and }\quad
\rho_{\rm cr} (1-\Omega _\Lambda
)u_0u_i\,,\label{160}\end{equation} which in the Newtonian limit
take the form
\begin{equation}{\rho_{\rm cr} c^2\over2}\,(1-3\Omega _\Lambda )\,,\qquad \mbox{and
}\quad \rho_{\rm cr} cv_i(1-\Omega
_\Lambda)\,,\label{170}\end{equation} where $\rho _{\rm
cr}=3H_0^2/8\pi G$; one term is multiplied by $(1-3\Omega
_\Lambda)$, the other by \mbox{$(1-\Omega _\Lambda)$}.  Note that
this neglects terms in derivatives of $\Omega _\Lambda$, but this is
acceptable since they contain the factor $H_0{\bf v}$ which is
always very small and, moreover, we take the limit $v/c\rightarrow
0$.

To obtain the result for the standard model of the universe with
$\Omega _\Lambda \neq 0$, one must use instead of (\ref{140}) the
expression
\begin{equation}{\bf g}^*=-{16\pi G\over c^2}\,\dot{{\bf v}}\,(1-\Omega _\Lambda )
\int _0^{R_U}\rho_{\rm cr}R{\rm d}R  + {4\pi G\over 3c^2}\,
\dot{\bf v}\,(1-3\Omega _\Lambda )\,\int _0^{R_U}\rho _{\rm cr}
R{\rm d}R \,, \label{175}
\end{equation} from which it follows easily that the
prediction for the reactive acceleration of the test mass due to the
rest of the universe is
\begin{equation} {\bf g}^*= {-11+9 \Omega
_\Lambda\over 4}\,(H_0t_0)^2 \dot{{\bf
v}}\,.\label{180}\end{equation} For $\Omega _\Lambda =0.7$, this is
$-1.09\,\dot{\bf v}$. It is a simple matter to show that the
reactive acceleration is between about $-1.2\,\dot{\rm v}$ and
$-1.1\,\dot{\rm v}$, according to eq. (\ref{180}), for all the
values of $\Omega _\Lambda$ in the interval $(0,\,0.9)$. This is
very close to eq. (\ref{10}). A warning however is necessary. The
reactive acceleration depends here on what we consider to be the
radius of the universe. Nevertheless, even taking into account the
simplifications in the physical model, this is intriguing (see
figure \ref{Figure 2}).

 \section{Second approach to the reactive acceleration}
 In this second approach, the integrals that give $h_{00}$ and $h_{0i}$
are calculated  along the past of our light cone,  using therefore
the retarded time. This solves the problem just mentioned of the
definition of the radius of the universe, although at the price of a
slight complication of the analysis.
  The equation of a light ray moving radially to or from Earth in the Robertson-Walker
 metric with $k=0$ (with constant $\theta$ and $\varphi$ and $\dif s=0$) is equal
 to
\begin{equation} c\,\dif t=\mp {S(t)\,{\rm d} r\over
(1-kr^2)^{1/2}}\label{190}\eeq where $S$ is the scale factor.

The double sign in (\ref{190}) is minus or plus according to whether
the photon aproaches to or recedes from Earth, respectively. Let us
take $k=0$. Denoting $f(t) =\int \dif t/S(t)$ and applying this
equation to an incoming photon that passes the radial coordinate $r$
at $t$ and reaches the Earth at $r=0$ and $t=t_0$, we have \beq
r=c\int _t^{t_0}{\dif t\over S}=c[f(t_0)-f(t)] \,.\label{200} \eeq
In the Einstein-de Sitter model (with $k=0$, $\Omega _M=1$,
$S=(t/t_0)^{2/3}$, $t_0=1/(6\pi G\rho _0)^{1/2}$ and $\rho (t)
=1/6\pi G t^2$ ), eq. (\ref{200}) gives $f(t)=3t_0^{2/3}t^{1/3}$ and
$r/c=f(t_0)-f(t)=3t_0(1-\tau^{1/3})$ with $\tau =t/t_0$. The proper
distance of the photon to us is therefore \beq
R=rS=3ct_0(\tau^{2/3}-\tau) \label{210}\,.\eeq This is the equation
of  the light cone at the cosmological level, represented in figure
\ref{Figure 1} as the solid line \cite{Rai01}. The line gives the
time as a function of the proper distance to the test mass along a
light ray with trajectory (\ref{210}). As we see, it has not shape
of cone but of onion. It could be called therefore ``the light
onion".
\begin{figure}[ht]
\begin{center}
\scalebox{0.9}{\includegraphics{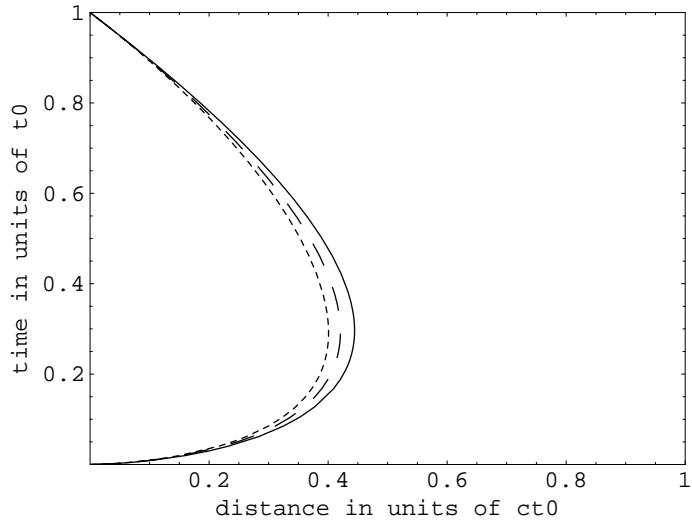}}
\end{center}
\caption{Light cones in the universe for the cosmological models
with $k=0$ and $\Omega _\Lambda =0$ (solid line), $\Omega_\Lambda  =
0.7$ (long dashes) and  $\Omega _\Lambda =0.9$ (short dashes). Each
curve referes to the particular value of $t_0$ corresponding to its
$\Omega _\Lambda$. The maximum value of the distance for $\Omega
_\Lambda =0$ is $R_{\rm max}=4ct_0/9$ at time $8t_0/27$ (explanation
in the text.)} \label{Figure 1}
\end{figure}
Light from a galaxy reaching us at present time is emitted at the
point of intersection of our past light-onion with the world line of
the galaxy. The maximum proper distance to the galaxy is $R_{\rm
max}=4ct_0/9$ at $\tau=8/27\simeq 0.296$. Note that for $\tau >8/27$
the photons approach Earth while for $\tau <8/27$ they recede from
it.

In the standard model, the integrals must be calculated over the
past light cone, {\it i. e.} with the distance $R$ and the time $t$
related by the curves in figure 1, each one obtained by inserting in
(\ref{200}), the corresponding scale factor $S(t)=((1-\Omega
_\Lambda)/\Omega _\Lambda)^{1/3}\sinh ^{2/3}[\sqrt{3\Lambda}\,t/2]$
with $\Lambda =3H_0^2\Omega _\Lambda$. For simplicity, we
approximate the different light cones for $0\le \Omega _\Lambda\leq
0.9$ by that of $\Omega _\Lambda =0$, which has a much simpler
analytical expression. This is justified because of their closeness.
In other words, we assume that the relation between $R$ and $t$ is
given by (\ref{210}), even if the expressions    for the
cosmological quantities as the density, the age of the universe or
the Hubble constant are assumed to depend on time according to the
standard model based on  corresponding scale factor. As usual, the
subindex $\Lambda$ indicate here present value, while $\lambda$
stands for the time dependent functions.

In order to proceed with the calculation, we must insert in the
integrals (\ref{175}) the retarded values of all the time
dependent quantities, {\it i. e.} of the density $\rho _{\rm
cr}=3H^2/8\pi G=(\Lambda /8\pi G)\coth^2[At]$ and $\Omega _\lambda
= \tanh ^2[At]$ with $A=\sqrt{3\Lambda /4}$. Since $\Lambda
=3H_0^2\Omega _\Lambda$, then $\rho _{\rm cr} =(3H_0^2\Omega
_\Lambda /8\pi G)\coth^2[At]$ and $A=3H_0\Omega _\Lambda
^{1/2}/2$.

Putting inside the integral the time dependent critical density
and $\Omega _\Lambda$, we have instead of (\ref{175})
\begin{equation}{\bf g}^*=-{16\pi G\dot{\bf v}\over c^2}
\int _{lc}\rho_{\rm cr}(t') (1-\Omega _\lambda(t'))R{\rm d}R +
{4\pi G\dot{\bf v}\over 3c^2}\, \,\int _{lc} \rho _{\rm
cr}(t')(1-3\Omega _\lambda (t'))R{\rm d}R \,,\label{230}
\end{equation}
where $t'$ is the retarded time along the light cone as given by eq.
(\ref{210}), from which it follows that $R\dif R = 9(ct_0)^2
f(\tau)\dif \tau$ with \beq f(\tau )= \epsilon (\tau) [(2/3)\tau
^{1/3}-(5/3)\tau ^{2/3}+\tau]\label{240} \eeq where $\epsilon
(\tau)$ is $-1$ for $\tau
> 8/27$ and $+1$ for $\tau <8/27$ in order to take into account
the two possible signs in eq. (\ref{190}), according whether the
photons approach to or recede from Earth.

We define the dimensionless non negative integrals \beq J= \int
_0^1\coth ^2 (a\tau)f(\tau)\dif \tau  \,,\quad K=\int _0^1\coth
^2(a\tau)\Omega _\lambda (a\tau) f(\tau)\dif \tau \,,\label{250}\eeq
where $\Omega _\lambda = \tanh^2(a\tau)$ and $a=At_0=3(H_0t_0)\Omega
_\Lambda ^{1/2}/2$. The second is immediate. In fact, it is easy to
see that $K= \int _0^1f(\tau )\dif \tau$ and that
$K=(3ct_0)^{-2}[\int _0^{R_{\rm max}}-\int _{R_{\rm max}}^0]R\dif R
=(3ct_0)^{-2}R_{\rm max}^2=16/729$, since $R_{\rm max}=4ct_0/9$.

After a bit of algebra, the reactive acceleration (\ref{230}) can
be written then as \beq {\bf g}^*=-\Omega _\Lambda (H_0t_0)^2
\left[{99\over 2}\, J- {8\over 9}\right]\,\dot{\bf v}\,,
\label{260} \eeq  which is the expression for the reactive
acceleration in our second approach and the main result of this
work. The integral $J$ can be written as the sum of two terms \beq
J=J_1+J_2= \int _{8/27}^1\coth ^2(a\tau) f(\tau) \dif \tau +\int
_0^{8/27}\coth ^2(a\tau )f(\tau )\dif \tau \label{280} \eeq
Unfortunately, $J_2$ is unbounded because of the divergence of the
integrand  at $\tau =0$.

Changing the variable from $\tau$ to $u=a\tau$, the integrals
$J_1$ and $J_2$  can be written as \begin{eqnarray} J_1&=&-{2\over
3a^{4/3}}I_{1/3}+{5\over 3a^{5/3}}I_{2/3}-{1\over
a^2}I_1,\nonumber
\\J_2&=&{2\over
3a^{4/3}}I'_{1/3}-{5\over 3a^{5/3}}I'_{2/3}+{1\over
a^2}I'_1\nonumber,\label{282}
\end{eqnarray}
where $I_\lambda =\int _{8a/27}^a\coth^2(u)\,u^\lambda \dif u$ and
$I'_\lambda =\int _0^{8a/27}\coth^2(u)\,u^\lambda \dif u$. The
three integrals $I'_\lambda$  are singular because the integrands
diverge at $u=0$. As is easy to show, however, that the
singularities are eliminated by subtracting to the integrands
$u^{-5/3}$, $u^{-4/3}$ and $u^{-1}$, respectively, what leaves
regular series which vanish at $u=0$. In fact, the three
integrands are, respectively, equal to
 \begin{eqnarray}&&u^{-5/3}
+ 2u^{1/3}/3+u^{7/3}/15+O(u)^{13/3}, \nonumber \\
&&u^{-4/3}+ 2u^{2/3}/3+u^{8/3}/15 + O(u)^{14/3},\nonumber \\
&&u^{-1} + 2u/3+u^3/15 +O(u)^5\nonumber \end{eqnarray} After
eliminating the singularities in this way, $J$ is easily calculated
and can be inserted in eq. (\ref{260}).

The result of this second approach for the coefficient $\xi$ (so
that the reactive acceleration is ${\bf g}^*=-\xi \dot{\bf v}$) are
plotted as the lower curve in Figure 2.  As is seen, $\xi$ is in the
interval $(0.7,\,0.6)$, approximately, if $\Omega _\Lambda <0.9$.
For $\Omega _\Lambda =0.7$, $\xi \simeq 0.65$. We estimate that the
accuracy of these numbers is about $10\,\%$, but we have made no
attempt to
 refine them any further.

Instead of eliminating the singularities this way, we could try to
find a cut-off. It can then be shown easily that, if the integral
$J_2$ is cut-off at $\tau \simeq 0.18$, corresponding to $t\simeq
2.5\mbox{ Gy}$, one obtains the right result $\xi =1$ for $\Omega
_\Lambda =0.7$.

This is not a bad result; quite on the contrary it is very
encouraging. The reactive acceleration is of the order of
$-\dot{\bf v}$ and varies little with $\Omega _\Lambda$.

\begin{figure}[ht]
\begin{center}
\scalebox{0.9}{\includegraphics{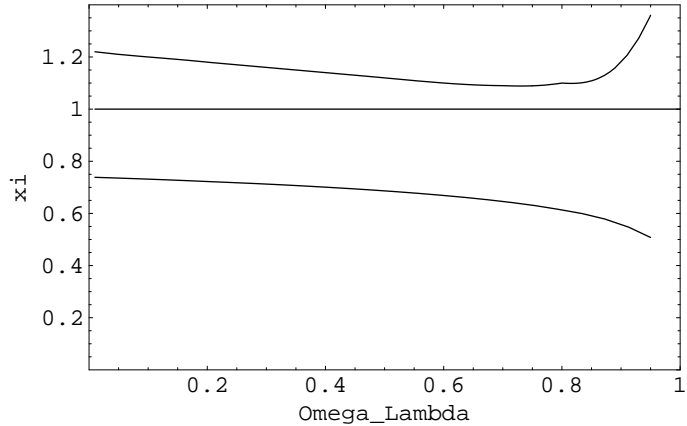}}
\end{center}
\caption{Coefficient $\xi$ in the reactive acceleration ${\bf
g}^*=-\xi \dot{\bf v}$ against $\Omega_ \Lambda$, in the first (eq.
(\ref{180}), upper line) and second (eq. (\ref{260}), lower line)
approaches, compared with the prediction of Mach's principle $\xi
=1$.} \label{Figure 2}
\end{figure}

\section{Summary and conclusions}
In order to evaluate Mach's principle, we have studied the
gravitational effect of the entire universe on an accelerated body
or test particle in its rest frame, using the formalism of
linearized gravitation. The aim is to show that the LHS of Newton
second law $\dot{\bf v}= {\bf F}/m$ is minus the reactive
acceleration due to the gravitational action of the all the
mass-energy in the universe on the particle in its rest frame. The
results are expressed in terms of two integrals depending on the
critical density and the relative density of dark energy. We have
followed two approaches. In the first one, the integrals are
calculated in the actual time section $t=t_0$ up to the radius of
the universe $R_U=ct_0$. In the second, the integration is done over
the past light cone of the particle, using the formalism of the
retarded fields. In both cases the results are enticing: the
reactive acceleration turns out to be close to $-\dot{\bf v}$: we
have found that, if $\Omega _\Lambda =0.7$, it is about
$-1.1\,\dot{\bf v}$ in the first approach, and close to
$-0.7\,\dot{\bf v}$ in the second.

All this means that the second Newton equation for a particle can be
obtained either as an approximation to the equation of the geodesics
or as the expression of the equilibrium between the forces due to
nearby objects and those of background gravity in the rest frame of
the particle. To summarize, our results strongly support Mach's idea
that inertia is not an intrinsic properties of bodies or particles
but an interaction, whose source fixes lapse $N$ and shift $N_i$,
and, furthermore, that this interaction is gravity as suggested by
Sciama.



\begin{thebibliography}{12345}

\bibitem{Mac72} E. Mach, {\it History and Root of the Principle of
Conservation of Energy} (Open Court Pub. La Salle, Illinois,
1911).
\bibitem{Mac83} E. Mach, {\it Die Mechanik in Ihrer Entwicklung: Historisch Kritisch
Dargestellt} (Brockhaus, Leipzig, 1883); English translation {\it
The Science of Mechanics: A Critical and Historical Account of its
development}, (6th ed., Open Court Pub., La Salle, Illinois,
1960).

\bibitem{See60} C. Seelig, {\it Albert Einstein} (Europa Verlag,
Z\"urich, 1960).

\bibitem{Ein16} A. Einstein, {\it Physicalische Zeitschrift}, {\bf
17}, 101 (1916).


\bibitem{Pai82} A. Pais, {\it Subtle is the Lord. The Life and
Science of Albert Einstein} (Oxford University Press, Oxford, 1982),
section 15e.

\bibitem{Ber89} M. V. Berry, {\it Principles of Cosmology and
Gravitation} (Institute of Physics Publishing, Bristol, 1989).

\bibitem{Sci53} D. W. Sciama, ``On
the origin of Inertia", {\it Monthly Notices of the Royal
Astronomical Society} {\bf 113}, 34 (1953).

\bibitem{Sci57} D. W. Sciama, ``Inertia", {\it Scientific American} {\bf 196},
no. 2, 99-109 (1957).

\bibitem{Sci69} D. W. Sciama, {\it The Physical Foundations of
General Relativity} (Doubleday $\&$ Co. New York, 1969).

\bibitem{Sci71} D. W. Sciama, {\it Modern Cosmology} (Cambridge
Univ. Press, Cambridge, 1971).

\bibitem{DeW64} B. S. Dewitt, {\it Phys. Rev.Lett.}  {\bf 13}, 114
(1964).
\bibitem{Alv89} E. \'Alvarez, {\it Rev. Mod. Phys.}  {\bf 61} 3, 561
(1989).

\bibitem{Mis73}  C. W. Misner, K. S. Thorne and J. A. Wheeler, {\it Gravitation}
(Freeman, San Francisco, 1973).

\bibitem{Rin01} W. Rindler, {\it Relativity. Spacial. General and
Cosmological} (Oxford University Press, Oxford, 2001).


\bibitem{Jac98} J. D. Jackson, {\it Classical Electrodynamics}, 3rd
edition (John Wiley, New York, 1998), chapter 14.

\bibitem{Bel81} Ll. Bel, Th. Damour, N. Deruelle, J. Ib\'a\~nez and J.
Mart\'in,
``Poincar\'e-Invariant Gravitational Field and Equations of Motion
of Two Point-like Objects: The Postlinear Approximation of General
Relativity", Gen. Rel. Grav. {\bf 13}, 963-1004 (1981).

\bibitem{Rai01} D. J. Raine and E. G. Thomas, {\it An introduction
to the science of cosmology} (IoP Publishing, Bristol, 2001),
chapter 5.


\end{thebibliography}
\end{document}